\documentclass[]{spie}  

\usepackage{amsmath,amsfonts,amssymb}
\usepackage{graphicx}
\usepackage[colorlinks=true, allcolors=blue]{hyperref}
\graphicspath{ {images/} }
\usepackage[skip=10pt]{caption} 
\usepackage{subcaption}
\usepackage{adjustbox}
\usepackage{pdfpages}
\usepackage{placeins}

\usepackage[utf8]{inputenc}

\usepackage{float}

\title{	Implementation of a broadband focal plane estimator for high-contrast dark zones}
\author[a]{Susan F. Redmond}
\author[b]{Laurent Pueyo}
\author[c]{Leonid Pogorelyuk}
\author[b]{James Noss}
\author[b,d]{Scott D. Will}
\author[e,f]{Iva Laginja}
\author[a,g]{N. Jeremy Kasdin}
\author[b]{Marshall D. Perrin}
\author[b]{Remi Soummer}

\affil[a]{Princeton University, Olden Street, Princeton, NJ 08544, USA}
\affil[b]{Space Telescope Science Institute, 3700 San Martin Drive, Baltimore, MD 21218, USA}
\affil[c]{Massachusetts Institute of Technology,77 Massachusetts Ave, Cambridge, MA 02139, USA}
\affil[d]{University of Rochester, 500 Joseph C. Wilson Blvd., Rochester, NY 14627, USA}
\affil[e]{DOTA, ONERA, Université Paris Saclay, F-92322 Châtillon, France}
\affil[f]{Aix-Marseille Université, CNRS, LAM (Laboratoire d’Astrophysique de Marseille) UMR 7326, 13388 Marseille, France}
\affil[g]{University of San Francisco, 2130 Fulton St, San Francisco, CA 94117, USA}

\authorinfo{Further author information: (Send correspondence to S.F.R.)\\S.F.R.: E-mail: sredmond@princeton.edu}

\begin{document} 
\maketitle

\begin{abstract}
The characterization of exoplanet atmospheres using direct imaging spectroscopy requires high-contrast over a wide wavelength range.  We study a recently proposed focal plane wavefront estimation algorithm that exclusively uses broadband images to estimate the electric field. This approach therefore reduces the complexity and observational overheads compared to traditional single wavelength approaches. The electric field is estimated as an incoherent sum of monochromatic intensities with the pair-wise probing technique. This paper covers the detailed implementation of the algorithm and an application to the High-contrast Imager for Complex Aperture Telescopes (HiCAT) testbed with the goal to compare the  performance between the broadband and traditional narrowband filter approaches.
\end{abstract}


\section{INTRODUCTION.}
\label{sec:intro}  
\subsection{Direct Observation and Spectroscopy of Exoplanets with a Space Telescope}

Due to the large brightness difference between a host star and its exoplanet, the exoplanet is often obscured by the lobes of the stars point spread function (PSF).  To remedy this, masks have been developed to suppress the starlight in regions around the star where planets of interest might exist; these regions are known as dark zones (DZ) or dark holes.  Masks alone are not sufficient, however, to observe dim planets at small inner working angles (IWA) from the host star.  Ground-based experiments have demonstrated the ability to suppress the starlight in the dark zone by a factor of $10^7$\cite{mesa_upper_2017}$^,$\cite{vigan_high-contrast_2015} by using fast deformable mirrors (DMs) to correct for the atmospheric turbulence.  To suppress the starlight by a factor of $10^{10}$ at small IWA (to observe an exo-Earth\cite{kasting_exoplanet_2009}), DMs are required to account for the manufacturing and alignment errors and a space telescope\cite{pueyo_luvoir_2019}$^,$\cite{krist_numerical_2016} is required to avoid the atmospheric effects that limit ground-based experiments.  The level of suppression is defined as the contrast and is expressed as
\begin{align}
    \text{contrast} = \frac{I_{m}}{\max{(I_{d})}}
    \label{eq:contrast}
\end{align}
where $I_{m}$ is the intensity in the masked, or coronagraph, image at the planet location and $I_{d}$ is the peak intensity in the un-masked, or direct, PSF.  It should be noted that at contrasts of $10^{-10}$, the electric field at the science instrument must be accurately estimated.  This means that focal plane wavefront estimation (FPWE) techniques must be used as pupil plane wavefront estimation techniques introduce non-common path errors\cite{sivaramakrishnan_sensing_2008}.   

Directly imaging exoplanets provides much more information about planetary systems than other complementary exoplanet detection techniques, such as transit detection\cite{rosenblatt_two-color_1971} or radial velocity\cite{wright_exoplanet_2013}.  One main advantage is the ability to do spectroscopy on the exoplanet\cite{wang_high_2021}$^,$\cite{danielski_atmospheric_2018} to determine its atmospheric composition.  To do spectroscopy on directly observed exoplanets, the starlight must be suppressed over a broad band of wavelengths\cite{trauger_hybrid_2011}.  In order to suppress the starlight within a certain band, an estimate of the electric field over the band must be obtained which can then be used to determine a DM command to reduce the light in the dark zone.  There are currently two accepted approaches to estimate the electric field within a broadband dark zone: (1) use an integral field spectrometer (IFS)\cite{sun_high-contrast_2020} and (2) use multiple narrowband filters\cite{groff_broadband_2012}.  For both the IFS and narrowband filter approaches, any of the monochromatic electric field estimators developed, such as pairwise-probing or Kalman filters\cite{groff_methods_2015}, can be used.   

For the IFS, a lenslet array combined with a prism is used to sample and disperse the incoming broadband light via wavelength.  Software (e.g.: \texttt{crispy}\cite{rizzo_simulating_2017}) is used to reconstruct monochromatic dark zone images at selected wavelengths within the band from the dispersed spectra.  The monochromatic dark zone images are then used to estimate the electric field at the selected wavelengths.  
While the IFS removes the need for narrowband filters, longer exposure times are required as the lenslet-pinhole array reduces the throughput\cite{chamberlin_integral_2016}.  The IFS also still requires additional hardware with tight alignment tolerances.

Using narrowband filters (monochromatic images) to estimate the electric field at the filter wavelengths is the most common approach to obtain an electric field estimate across a broad band.  The quality of the broadband dark zone is dependent on how many filters are used.  Using narrowband filters for broadband estimation is relatively simple and does not introduce strict alignment requirements other than the planarity of the filters\cite{groff_broadband_2012}.  The narrowband filter approach does require a factor of $l$ more images than the IFS approach where $l$ is the number of filters.   

In this paper we discuss a third option for estimating the electric field in a broadband dark zone, the Broadband estimator, that attempts to combine the benefits of both the IFS and narrowband filter approaches.  The Broadband estimator, as discussed in Pogorelyuk et al. 2020 \cite{pogorelyuk_effects_2020}, uses the pairwise-probing method but uses broadband images instead of monochromatic ones.  The observation matrix is also broadband, containing $l$ Jacobians, and sums the contributions from each wavelength for each pixel.  Initial results are obtained using the High-contrast Imager for Complex Aperture Telescopes (HiCAT) at the Space Telescope Science Institute (STScI).  All results presented in this paper are hardware results, but HiCAT has a high-fidelity simulator which emulates the testbed (\texttt{catkit}\cite{noss_spacetelescopecatkit_2021}) and is extensively used for developing algorithms prior to hardware tests.  HiCAT contains a classical Lyot coronagraph, two Boston Micromachine (BMC) kilo DMs, and an IrisAO PTT111L segmented aperture.  The current broadband source is a Leukos SM-30-400 laser combined with a filter wheel.  The filter wheel has five 10~nm bandpass filters available at 610, 620, 640, 660, and 670~nm as well as a 6\% bandpass filter centered at 640~nm.  The initial results from HiCAT using broadband images for FPWE combined with an Electric Field Conjucation (EFC) controller show a reduction of the intensity in the dark zone by an order of magnitude when the 6\% bandpass filter is used and six wavelengths within the band are estimated.  The Broadband estimator is still in the early stages of development and more work is required to improve the performance and determine if it is a viable option for on-sky operations.

\section{Broadband Estimator Algorithm.}\label{sec:bb_est_alg}

The Broadband estimator is an extension of the monochromatic Pairwise-probe estimator\cite{groff_methods_2015} commonly used to generate a dark zone.  DM probe commands are developed to modulate the electric field at the focal plane and images are taken using the positive and negative versions of those DM commands.  By taking the difference of the positive and negative probed images we can estimate the electric field.  The main difference for the Broadband estimator is that the electric field for multiple wavelengths is estimated simultaneously.  To do this, we must assemble a broadband Jacobian containing information for all $l$ wavelengths of interest.  The monochromatic Jacobians are calculated as    
\begin{align}
    \mathbf{G}_{i} = \sqrt{t_{exp}}\begin{bmatrix}
        \Re\left(\frac{\partial E_0}{\partial u}\right)\\\\
        \Im\left(\frac{\partial E_0}{\partial u}\right) \\
        \vdots\\
        \Re\left(\frac{\partial E_n}{\partial u}\right)\\\\
        \Im\left(\frac{\partial E_n}{\partial u}\right)
    \end{bmatrix}
\end{align}
by taking the partial derivative of the electric field ($E$) at each pixel (0--$n$) in the dark zone with respect to each DM actuator for wavelength $i$.  The estimator operates in units of counts so the Jacobian must be multiplied by the square root of the exposure time ($t_{exp}$) every iteration.  To have a purely real Jacobian, the real and imaginary components of the partial derivatives are stacked in a pixel-wise manner.  The monochromatic Jacobians are merged to produce a broadband Jacobian defined as  
\begin{align}
    \mathbf{G}_{bb} = \begin{bmatrix}
        \mathbf{G}_{0}(0,:)\\
        \mathbf{G}_{0}(1,:)\\
        \mathbf{G}_{1}(0,:)\\
        \mathbf{G}_{1}(1,:)\\
        \vdots\\
        \mathbf{G}_{l-1}(0,:)\\
        \mathbf{G}_{l-1}(1,:)\\
        \vdots\\
        \mathbf{G}_{l-1}(n-2,:)\\
        \mathbf{G}_{l-1}(n-1,:)
    \end{bmatrix}
\end{align}
where $l$ is the number of wavelengths.  This groups the information for each pixel since $\mathbf{G}_{i}(k,:)$ and $\mathbf{G}_{i}(k+1,:)$ are the real and imaginary components for the same pixel.  In general, assuming DM commands are small, the monochromatic intensity at the focal plane after a DM command is applied can be approximated as
\begin{align}
    I_i =  \mathbf{B}_i|E_i + \mathbf{G}_iu|^2
\end{align}
where $E$ is the electric field prior to the command being applied, $I$ is the intensity at the focal plane after the command is applied, $i$ is the wavelength, and $u$ is the DM command.  In the monochromatic case, $\mathbf{B}_i$ is a matrix that sums the squared real and imaginary components electric field.  Note that the electric field is vector is split in a similar manner to the Jacobians as shown by
\begin{align}
    E_i =  \begin{bmatrix}
        \Re(E_{p=0})\\
        \Im(E_{p=0})\\
        \vdots\\
        \Re(E_{p=n})\\
        \Im(E_{p=n})\\
    \end{bmatrix}
\end{align}
where $p$ is the pixel.  Using the small command assumption if we subtract the images at wavelength $i$ obtained from a positive ($I_{j+}$) and negative ($I_{j-}$) probe command $u_j$ we then get
\begin{align}
    I_{i,j+} - I_{i,j-} &= \mathbf{B}_i|E_i + \mathbf{G}_iu_j|^2 - \mathbf{B}_i|E_i - \mathbf{G}_iu_j|^2 = 4\mathbf{B}_i\text{diag}(\mathbf{G}_iu_j)E_i \\
    E_i &= \mathbf{H}_i\left(I_{i,j+} - I_{i,j-}\right)
\end{align}
which is the monochromatic pairwise-probe estimator where $\mathbf{H} = \left(4\mathbf{B}_i\text{diag}(\mathbf{G}u_j)\right)^+$.  Note that to make $\mathbf{H}$ well conditioned, multiple sets of probed images must be obtained.  Even with multiple sets of probes, a regularization parameter is often required to perform the psuedo-inverse.  When this is expanded to create the Broadband estimator, broadband images are taken and $\mathbf{B}$ now sums the wavelength components of $\mathbf{G}_{bb}u_j$ for each pixel as well as the real and imaginary components. This produces the broadband observation matrix
\begin{align}
    \mathbf{H} = \begin{bmatrix}
        4\mathbf{B}\cdot\text{diag}(\mathbf{G}_{bb}u_0)\\
        \vdots\\
        4\mathbf{B}\cdot\text{diag}(\mathbf{G}_{bb}u_s)
    \end{bmatrix}^+
\end{align}
where $s$ is the number of probes.  The observation matrix is then used to determine the electric field at the selected wavelengths,
\begin{align}
    \begin{bmatrix}
        \Re(E_{0,0})\\
        \Im(E_{0,0})\\
        \Re(E_{0,1})\\
        \Im(E_{0,1})\\
        \vdots\\
        \Re(E_{0,l})\\
        \Im(E_{0,l})\\
        \Re(E_{1,0})\\
        \Im(E_{1,0})\\
        \vdots\\
        \Re(E_{n,l})\\
        \Im(E_{n,l})\\
    \end{bmatrix} = \mathbf{H} \begin{bmatrix}
        I_{bb,0+} - I_{bb,0-}\\
        \vdots\\
        I_{bb,s+} - I_{bb,s-}
    \end{bmatrix} 
    \label{eq:bb_est}
\end{align}
where $n$ is the number of pixels, $l$ is the number of wavelengths, and all the images are broadband images.  To check the accuracy of the estimate, the broadband intensity estimate is calculated and compared to the un-probed broadband image.  The estimator operates in units of counts but we prefer the error in units of contrast so the broadband intensity estimate ($\hat{I}_{bb}$) is calculated as   
\begin{align}
    \hat{I}_{bb} = \sum_{i=1}^{l} \frac{\Re(E_{\lambda_i})^2 + \Im(E_{\lambda_i})^2}{\max{I_{d_{bb}}}}
\end{align}
where $\max{I_{d_{bb}}}$ is the peak intensity of the un-masked broadband PSF as described in Eqn.~\eqref{eq:contrast}.  The estimate error is then expressed as 
\begin{align}
    \epsilon_{bb} = |I_{bb} - \hat{I}_{bb} |
    \label{eq:est_err}
\end{align}
and will be discussed more in Section \ref{sec:results}.

\begin{figure} [htb]
	\centering
   \includegraphics[scale=0.45]{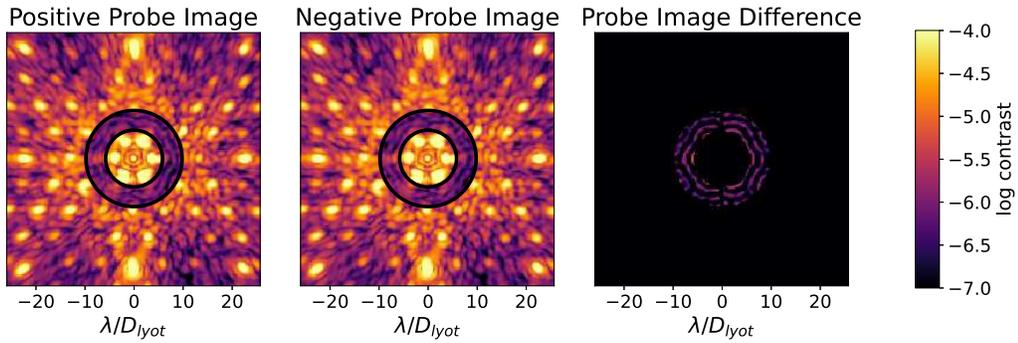}
   \caption[Probe sample] 
   {\label{fig:probe_sample} Example probe for the Broadband estimator ($\theta=0$).  The difference between the positive and negative probe images cannot be seen by eye but the phase introduced is easily seen in the probe image difference image in the right panel.  The probe images use the 6\% filter as is evident by the smearing of the satellite spots towards the edge of the frame.}
\end{figure}

The DM probe commands are an important aspect of the Broadband estimator.  For this paper, `ordinary' probes are used which introduce a uniform phase across the dark zone, determined by $\theta$, and are calculated via
\begin{align}
    u_j = (\mathbf{G}_i^T \mathbf{G}_i + 
    \alpha_{probe} \mathbf{I})^{-1}\mathbf{G}_i^T  \begin{bmatrix}
        \text{cos}~\theta_j\\
        \text{sin}~\theta_j\\
        \vdots
    \end{bmatrix} 
\end{align}
where $\alpha_{probe}$ is the regularization parameter and cos/sin alternate for all $n$ pixels in the dark zone.  For a given set of $s$ probes, $\theta$ sweeps from 0--$\pi$ in equal steps.  A sample broadband probe for $\theta_0=0$ is shown in Fig.~\ref{fig:probe_sample} for iteration 18 of a focal plane wavefront control (FPWC) experiment. The amplitude of the probe is updated each iteration based on the contrast; a higher contrast requires a smaller probe amplitude.  It is important to note that the probes are wavelength dependent.  Currently, the probes are calculated using the Jacobians for the centre wavelength of the band.  Further investigation into improving the probe effectiveness is in progress.    

Once an estimate is obtained for each of the desired wavelengths, Electric Field Conjugation (EFC)\cite{giveon_broadband_2007} is used to determine the optimal DM command.  As discussed in Pogorelyuk et al. 2020\cite{pogorelyuk_effects_2020}
, this improves the numerical conditioning of the overall algorithm.  The DM command based on a single wavelength estimate is expressed as
\begin{align}
    \Delta u_i = (\mathbf{G}_i^T \mathbf{G}_i + 
    \alpha \mathbf{I})^{-1}\mathbf{G}_i^T \hat{E}_i
\end{align}
where $\alpha$ is the Tikhonov regularization parameter that limits the maximum allowable DM command.  For broadband control, the command is taken as the mean of the EFC output for each wavelength,
\begin{align}
    \Delta u = \frac{1}{l}\sum_{i=0}^{i=l}\Delta u_i.
\end{align}
For future experiments we plan on implementing the broadband EFC algorithm discussed in Give'on et al. 2007\cite{giveon_broadband_2007}.

\section{Laboratory Results.}\label{sec:results}
Using a dark zone size of 5.8--9.8~$\lambda/D_{lyot}$ on HiCAT, there are six very bright lobes near the IWA.  These lobes can be difficult to suppress when they are on the border of the dark zone.  To improve suppression of the bright inner lobes, the dark zone is initially dug to a contrast of $10^{-6}$ using an oversized region with an IWA of 3.8~$\lambda/D_{lyot}$.  The DM command obtained by the oversized dark zone run is referred to as the `resume command' and is used as a starting point for the `real' experiments.  In some cases the resume command is obtained for only the 640~nm band which can cause issues for the Broadband estimator as the starting contrast will not be uniform across the wavelength band; this will be demonstrated in Section \ref{subsec:cube_mode}.  

\subsection{Cube Mode}\label{subsec:cube_mode}
As an intermediate step between monochromatic and true broadband estimation, the `cube mode' of the Broadband estimator was developed.  In cube mode, images are taken using 10~nm bandpass filters and added together (in intensity units, not contrast) to create a stand-in broadband image.  The estimator only has access to the broadband image but the monochromatic images can be used to check the estimate quality.  Cube mode would never be used for on-sky operations. 

\begin{figure} [htb]
	\centering
   \includegraphics[scale=0.4]{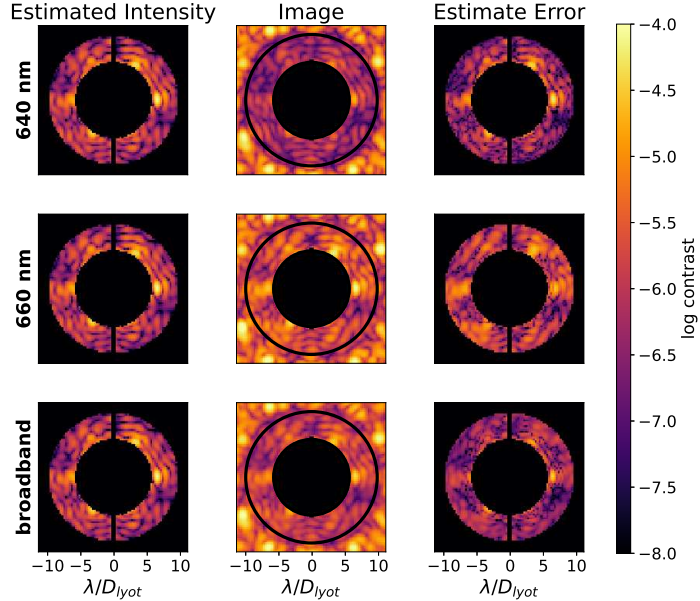}
   \caption[Cube Mode Estimate Comparison] 
   {\label{fig:cube_mode_estimate} Comparison of the estimates and images for iteration 0 in the cube mode experiment.  The bottom row contains broadband data which is calculated via Eqn.~\eqref{eq:bb_contrast}.  The left column contains the estimates where the top two rows are calculated by the estimator.  The middle column contains the images; the bottom broadband image is the type of image that is provided to the estimator.  The center of the images is masked to highlight the features in the dark zone near the IWA.  Here we can see the effect of using the resume DM command as the 640~nm dark zone starts off with a much higher contrast.  Since all estimated wavelengths are assumed to contribute equally to the broadband intensity, the electric field is over-estimated at 640~nm and under-estimated at 660~nm.  The right column contains the estimate error described by Equation~\eqref{eq:est_err} where we can see that the 660~nm estimate error is much larger than the 640~nm estimate error, again, due to the use of the resume command.  }
\end{figure}

\begin{figure}[htb]
	\centering
 	\begin{subfigure}[b]{0.5\textwidth}
 	    \centering
    	\includegraphics[scale=0.47]{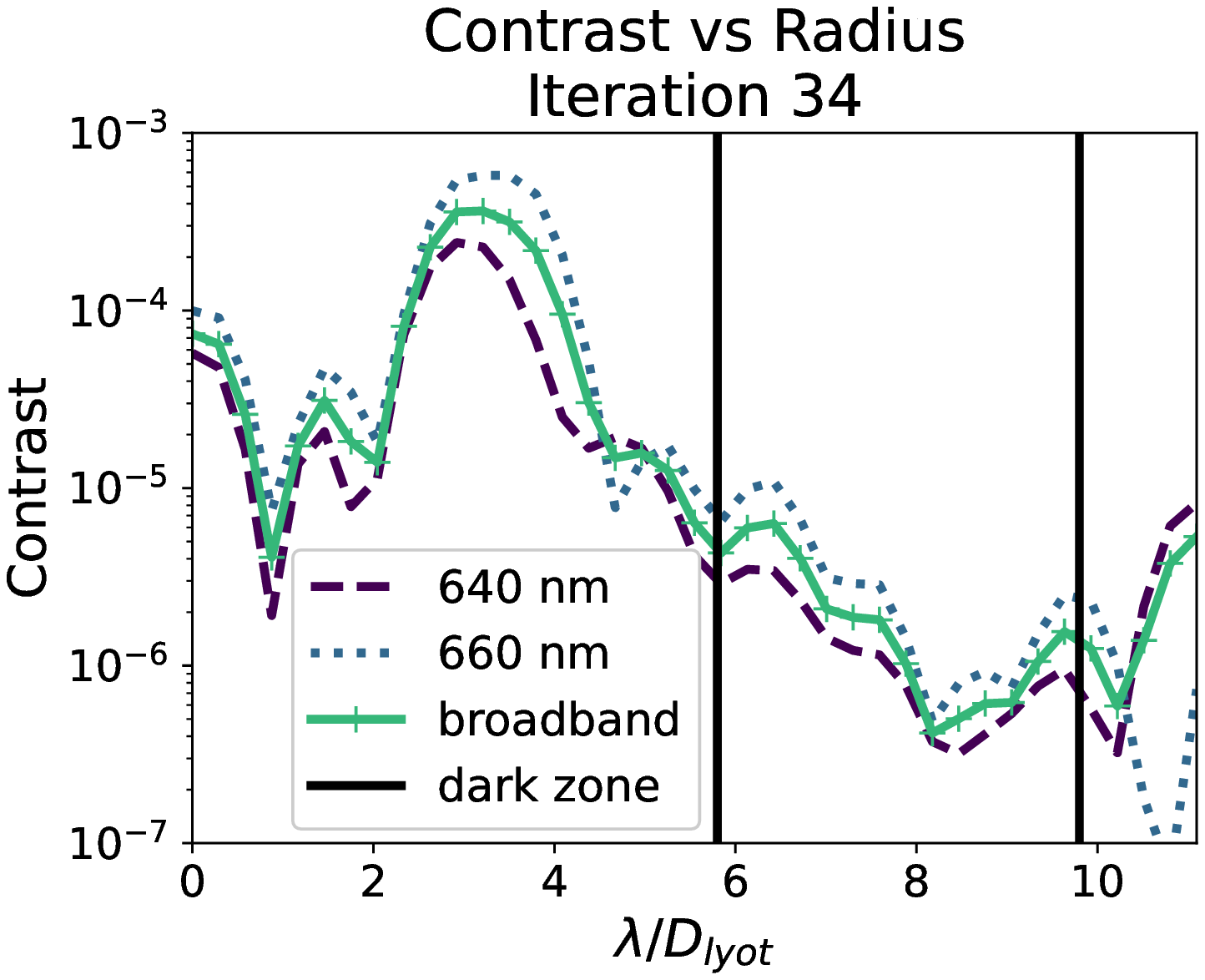}
    	\captionsetup{justification=centering}
	    \caption{Contrast vs radius for iteration 34, t=2~hrs}
	    \label{fig:cont_vs_rad_cube}
 	\end{subfigure}
 	\begin{subfigure}[b]{0.45\textwidth}
 		\centering
    	\includegraphics[scale=0.47]{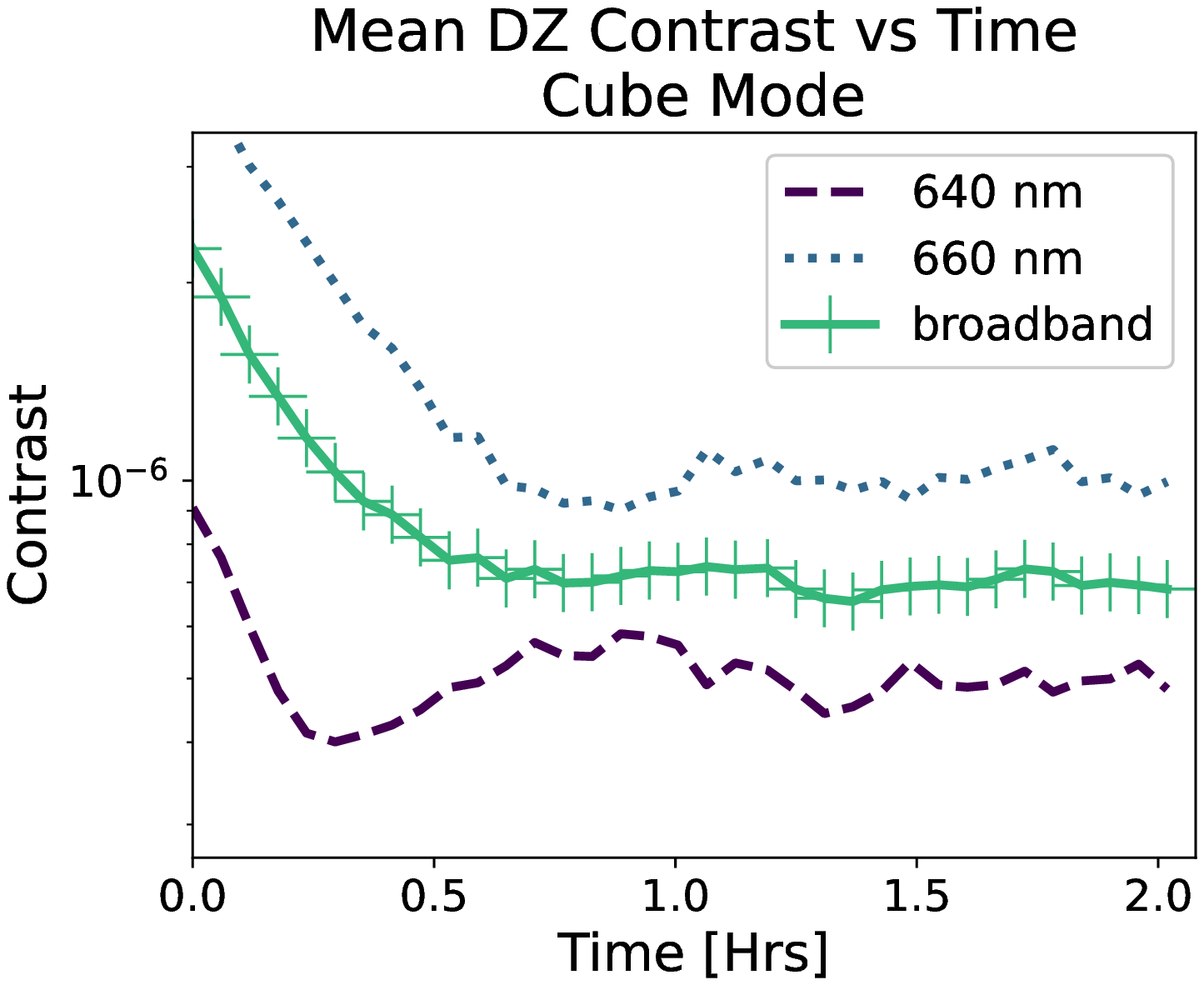}
    	\captionsetup{justification=centering}
	    \caption{Mean dark zone contrast vs time for cube mode.}
	    \label{fig:cont_vs_time_cube}
 	\end{subfigure}
 	\caption{Contrast plots for cube mode experiment.  Figure~\ref{fig:cont_vs_rad_cube} shows the worst case scenario contrast vs. radius with an IWA broadband contrast of $\sim3.5\times10^{-6}$ and an OWA broadband contrast of $\sim2\times10^{-6}$. 
 	Figure~\ref{fig:cont_vs_time_cube} shows the mean dark zone contrast at each wavelength as well as the broadband mean dark zone contrast vs. time.  The 640~nm contrast begins at a factor of four better contrast than the 660~nm due to the resume command used.  By iteration 34 (2~hrs), the monochromatic contrasts are within a factor of two.  The steady state broadband contrast is $6.9\times10^{-6}$.}
 	\label{fig:cont_vs_cube}
 \end{figure}

To demonstrate cube mode, Fig. \ref{fig:cube_mode_estimate} and \ref{fig:cont_vs_cube} show results from an experiment run using the 640 and 660~nm filters.  The 640 and 660~nm filters are chosen as they provide a 20~nm band that span a 30~nm band (5\%) and should provide a good reference point for when the 6\% filter is used.  For this experiment, the estimating wavelengths are also 640 and 660~nm.  The initial DM command is a 640~nm resume command (as described above).  Six probed images are taken at each wavelength and added together to obtain $I_{bb,+}, I_{bb,-}$ (to be used in Eqn.~\eqref{eq:bb_est}) for each of the $s$ probes.  After the monochromatic electric field estimates are obtained using the probed images, the broadband estimate error is calculated via Eqn.~\eqref{eq:est_err}.

As shown in Fig. \ref{fig:cube_mode_estimate}, the Broadband estimator has moderate success at estimating the electric field at the selected wavelengths.  The left column shows the estimates at each wavelength, the center column shows the images at each wavelength, and the right column shows the error between the left and center columns.  Note that the contrast for the first two rows is expressed as $I_{m_i} / \max{I_{d_i}}$, where $i$ is the wavelength, but the bottom row contrast is expressed as
\begin{align}
    \text{contrast}_{bb} = \sum_{i=1}^l \frac{I_{m_i}}{ \max{I_{d_{bb}}}}\label{eq:bb_contrast}
\end{align}
as the peak intensity of the stand-in direct broadband image ($I_{d_{bb}}$) is approximately twice that of the individual wavelengths.  The effect of using the 640~nm resume command is evident in the middle column as the initial 640~nm dark zone is much more pronounced.  The estimator assumes all wavelengths to be equal contributors to the broadband intensity and thus over-estimates the 640~nm electric field and under-estimates the 660~nm electric field. 

 Figure~\ref{fig:cont_vs_cube} shows the contrast plots for the cube experiment.  Note that the \ref{fig:cont_vs_rad_cube} is a worst case scenario as it is taken along the $y=0$ line where the residual spoke pattern seen in Fig.~\ref{fig:cube_mode_estimate} is most prominent.
Figure~\ref{fig:cont_vs_time_cube} shows the broadband and monochromatic mean dark zone contrast vs. time. Note that the mean dark zone contrast refers to the spatially averaged contrast within the 5.8--9.8~$\lambda/D_{lyot}$ annulus.  Here we can see that the 640~nm resume command causes the initial 640~nm contrast (purple dashes) to be a factor of four better and the steady state contrast to be a factor of two better than the 660~nm contrast (blue dots).  The large contrast discrepancy between the two wavelengths is likely what causes the 640~nm contrast increase after t=0.25~hrs.

\subsection{Broadband Mode}\label{subsec:bb_mode}

\begin{figure} [htb]
	\centering
   \includegraphics[scale=0.28]{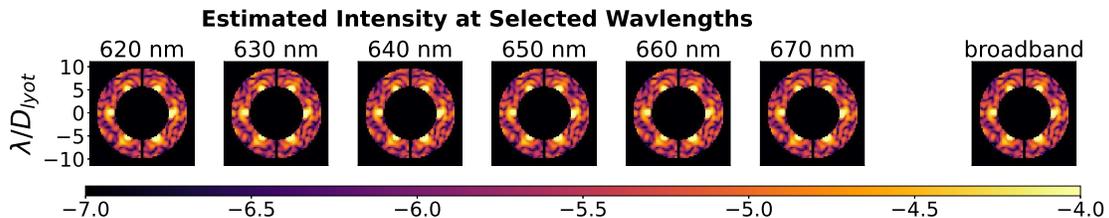}
   \caption[Broadband estimates] 
   {\label{fig:bb_est_breakdown} Broadband estimator estimate breakdown for iteration 0.}
\end{figure}

Broadband mode is the approach that would be used on-sky where a single broad filter is used to acquire the broadband probe images.  On HiCAT we use a 6\% filter centered at 640~nm.  The broadband probe images are then plugged into Eqn.~\eqref{eq:bb_est} to estimate the electric field at the chosen wavelengths.  For this experiment, increments of 10~nm were chosen from 620--670~nm as the estimation wavelengths to match the filters available to the narrowband filter approach.  Also, the transmission of the filter is $>$90\% from 616--662~nm and we wanted to ensure the edges of the band were covered.  In future experiments we will likely not estimate 670~nm when using the 6\% filter.  For this experiment, 30 probes were used which is five probes per estimated wavelength.  Figure~\ref{fig:bb_est_breakdown} shows the product of the Broadband estimator for iteration 0.  The right broadband panel is computed separately after.  As expected the estimates have similar features but are not identical and differ at the $10^{-7}$ level.       

\begin{figure} [htb]
	\centering
   \includegraphics[scale=0.45]{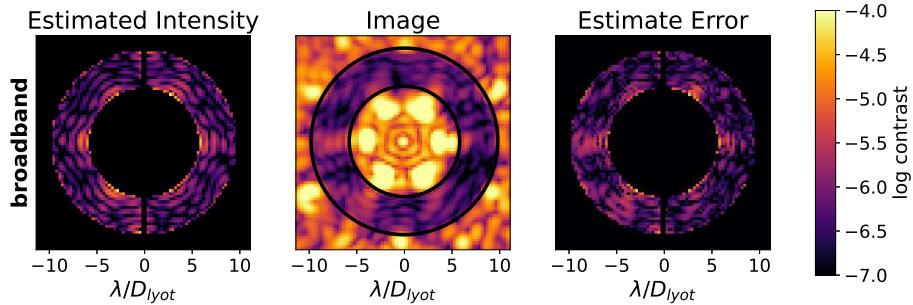}
   \caption[Broadband estimate comparison] 
   {\label{fig:bb_est_comp} Broadband estimate comparison for iteration 20.}
\end{figure}

Figure~\ref{fig:bb_est_comp} shows the comparison of the broadband estimate for iteration 20 and the broadband image.  This experiment does not use the resume command and the six bright lobes inside the IWA are very evident.  The estimator captures the hot spots at the IWA but does not capture the full spokes as they extend into the dark zone.    

\begin{figure} [htb]
	\centering
   \includegraphics[scale=0.55]{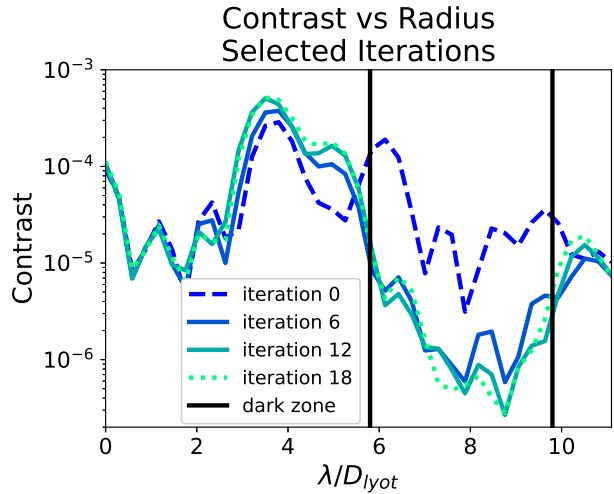}
   \caption[Broadband estimate comparison] 
   {\label{fig:bb_cvt} Broadband contrast vs radius for select iterations.  Iteration 0 is shown by the dark blue dashes and iteration 18 is shown by the green dots.  The dark zone limits are shown by the black vertical lines.  This slice is taken through the center of the dark zone where $y=0$ and is a worst case scenario since it contains the residual spoke pattern.  The bulk of the correction occurs in the first five iterations at which point the contrast begins to asymptote. }
\end{figure}

Figure~\ref{fig:bb_cvt} looks at the contrast vs. radius and how that varies with time.  The first iteration is shown by the dashed blue lines and the third last iteration is shown by the green dots.  The majority of the correction occurs in the first five iterations at which point the algorithm fails to continue removing the six-fold spoke pattern.  As in Section \ref{subsec:cube_mode}, the slice in Fig.~\ref{fig:bb_cvt} is along $y=0$ which is a worst case scenario since it follows one of the spokes.  It could be that we need to run a short experiment with the Broadband estimator using the 6\% filter and a smaller IWA to kill the bright lobes and the spoke pattern but it is our hope that we can achieve this by improving the estimator instead.    


\subsection{Mode Comparison}\label{subsec:mode_comp}
\begin{figure} [htb]
	\centering
   \includegraphics[scale=0.5]{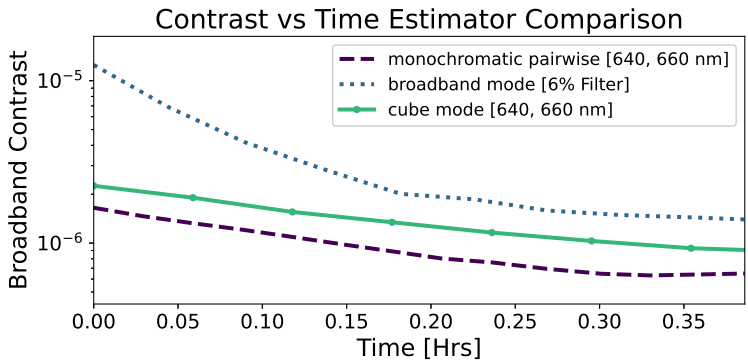}
   \caption[Dark hole generation comparison] 
   {\label{fig:bb_sm_cube_comp} Comparison of methods to generate a broadband dark zone on HiCAT.  The Broadband estimator approaches are shown by the blue dotted line and green solid line representing broadband mode and cube mode respectively.  The monochromatic pairwise approach is shown by the purple dashes.  The monochromatic pairwise method achieves the highest contrast but all curves have approximately the same slope once they hit $2\times10^{-6}$.}
\end{figure}

HiCAT has not yet been optimized for broadband performance so it is not quite fair to judge the absolute dark zone digging performance of the Broadband estimator.  In Fig.~\ref{fig:bb_sm_cube_comp} we compare different methods of generating a dark zone on HiCAT.  It should be noted that the cube mode (green solid curve) and monochromatic pairwise approach (purple dashes) utilize the resume command.  The cube mode and monochromatic pairwise method are both only estimating 640 and 660~nm as they tend to have stability issues when incorporating more wavelengths.  The stability issues can be addressed for the most part by choosing very conservative controller parameters but that results in very slow dark zone generation.  Broadband mode as shown by the blue dots in Fig.~\ref{fig:bb_sm_cube_comp} is slightly more robust and can have slightly more aggressive controller parameters.  Though the monochromatic pairwise method achieves the highest mean dark zone broadband contrast ($6.3\times10^{-7}$), the slopes of all of the curves are approximately equivalent after they pass $2\times10^{-6}$.  This is a good sign as it indicates the Broadband estimator performance is not significantly worse than its monochromatic partner.  It should also be noted that the monochromatic pairwise experiment used four probes per wavelength estimated, the cube mode experiment used three probes per wavelength estimated, and the broadband mode used five probes per wavelength estimated.          

\section{Conclusions and Future Work.}\label{sec:conclusions}

Generating a dark zone across a wide wavelength band is an important capability for a high contrast imaging system.  In order to minimize hardware requirements, we investigate the potential of using broadband images to estimate the electric field at discrete wavelengths within the band.  We demonstrate focal plane wavefront estimation and control using broadband images where the mean dark zone contrast is increased by an order of magnitude (from $10^{-5}$ to $10^{-6}$) across a 6\% band centered at 640~nm.  Using images taken with the 6\% filter we estimate the electric field at 10~nm increments from 620--670~nm.

Though we have demonstrated the potential of the Broadband estimator, there are many areas of improvement.  A common issue encountered throughout the experiments presented in this paper is the accuracy of the Jacobians.  Extensive model matching has only been done for the 640~nm Jacobian causing the contrast at 640~nm to improve much more quickly than the other wavelengths.  Performing model matching at other wavelengths is expected to improve the performance of both the Broadband estimator and the monochromatic pairwise estimator.  Moving forward a main point of interest is the probes used for the Broadband estimator.  In Pogorelyuk et al. 2020\cite{pogorelyuk_effects_2020} random probes were used but these were found to perform worse in the HiCAT simulator and on the HiCAT testbed.  The goal is to induce chromatic effects that probe the entire dark zone.  For this reason using Sinc probes on DM2 might be a better option than ordinary probes on DM1.  Lastly, the chromaticity of the testbed has not been considered in this paper.  Adding spectral weights to the estimated electric fields may also improve the Broadband estimator performance.

\acknowledgments 
 
This work was supported in part by the National Aeronautics and Space Administration under Grant\\ 80NSSC19K0120 issued through the Strategic Astrophysics Technology/Technology Demonstration for Exoplanet Missions Program (SAT-TDEM; PI: R. Soummer).

\bibliography{zotero_references} 
\bibliographystyle{spiebib} 

\end{document}